\documentclass{aa}
\usepackage[varg]{txfonts}

\usepackage{hyperref}
\hypersetup{colorlinks=true,linkcolor=blue,citecolor=blue,filecolor=blue,urlcolor=blue}

\usepackage{tikz}
\usepackage{graphicx}
\graphicspath{{images/}}

\usepackage{subfigure}

\usepackage{multicol}
\usepackage{multirow}
\usepackage{makecell, booktabs}
\usepackage{ulem}

\usepackage{bm}

\newcommand{\tid}{\mathrm{tid}}

\newcommand{\noleft}{\left.\kern-\nulldelimiterspace}

\newcommand{\Nsph}{N_\mathrm{sph}}
\newcommand{\Vrms}{V_\mathrm{rms}} 
\newcommand{\rhoc}{\rho_\mathrm{c}} 
\newcommand{\kkick}{k_\mathrm{kick}} 
\newcommand{\rvir}{r_\mathrm{vir}} 
\newcommand{\tcc}{t_\mathrm{cc}} 
\newcommand{\Xitid}{\Xi_\mathrm{tid}} 
\newcommand{\xcore}{\boldsymbol{x_\mathrm{c}}} 
\newcommand{\rcore}[1][]{r_\mathrm{c#1}} 
\newcommand{\Mlost}{M_\mathrm{lost}} 
\newcommand{\tauw}{\tau_\mathrm{w}} 

\newcommand{\kms}{\mathrm{km\,s^{-1}}}
\newcommand{\ccm}{\mathrm{cm^{-3}}}
\newcommand{\pc}{\mathrm{pc}}
\newcommand{\MSun}{M_\odot}

\newcommand{\Myr}{\mathrm{Myr}}

\definecolor{lime}{HTML}{A6CE39}
\DeclareRobustCommand{\orcidicon}{%
        \begin{tikzpicture}
        \draw[lime, fill=lime] (0,0) 
        circle [radius=0.16] 
        node[white] {{\fontfamily{qag}\selectfont \tiny ID}};
        \draw[white, fill=white] (-0.0625,0.095) 
        circle [radius=0.007];
        \end{tikzpicture}
        \hspace{-2mm}
}

\newcommand{\orcidPS}{\href{https://orcid.org/0000-0001-7044-3809}{\orcidicon}}
\newcommand{\orcidSS}{\href{https://orcid.org/0000-0003-1677-8004}{\orcidicon}}
\newcommand{\orcidVP}{\href{https://orcid.org/0000-0002-3031-062X}{\orcidicon}}

\begin{document}

\title{Environmental Effects on the Dynamical Evolution of Star Clusters in Turbulent Molecular Clouds}
\subtitle{}

\titlerunning{Star Clusters in Turbulent Molecular Clouds}

\author{Paolo Suin \inst{\ref{pisa}}\thanks{\email{p.suin@studenti.unipi.it}}\orcidPS
\and Steven N. Shore \inst{\ref{pisa},\ref{infn}}\orcidSS
\and V\'aclav Pavl\'ik \inst{\ref{iu}}\orcidVP}

\institute{Dipartimento di Fisica, Universit\`a di Pisa, largo B. Pontecorvo 3, Pisa 56127, Italy \label{pisa}
\and INFN, Sezione di Pisa, largo B. Pontecorvo 3, Pisa 56127, Italy \label{infn}
\and Indiana University, Department of Astronomy, Swain Hall West, 727 E 3$^\text{rd}$ Street, Bloomington, IN 47405, USA \label{iu}}

\authorrunning{Suin, Shore \& Pavl\'ik}

\date{Received 18 March 2022 / Accepted 1 July 2022}

\abstract
{Star clusters form within giant molecular clouds that are strongly altered by the feedback action of the massive stars, but the cluster still remains embedded in a dense, highly turbulent medium and interactions with ambient structures may  modify its dynamical evolution from that expected if it were isolated.}
{We aim to study coupling mechanisms between the dynamical evolution of the cluster, accelerated by the mass segregation process, with harassment effects caused by the gaseous environment.}
{We simulated the cluster dynamical evolution combining $N$-body and hydrodynamic codes within the Astronomical Multipurpose Software Environment (\texttt{AMUSE}).}
{Tidal harassment produces a sparser configuration more rapidly than the isolated reference simulations. The evolution of the asymptotic power-law density distribution exponent also shows substantially different behaviour in the two cases. The background is more effective on clusters in advanced stages of dynamical development.}
{}


\maketitle

\section{Introduction and physical scenario}
\label{sec:intro}

Understanding the evolution of star cluster has been the subject of many efforts over the past decades. Much work has been done to uncover how the internal effects of dynamical interactions \citep[e.g.,][]{  Spitzer1940, Chandrasekhar1942, Lynden-Bell1980, Spitzer1987} and stellar evolution \citep[e.g.,][]{Chernoff1990,PortegiesZwart2007} impact its properties. Yet no cluster is truly isolated, and modelling the interaction with the environment, such as the large scale Galactic potential, spiral arms and high-density structure in the interstellar medium \citep[ISM; e.g.,][]{ Spitzer1987, Terlevich1987, Gnedin1997, Gieles2006} also received great attention from the scientific community. Indeed, star clusters and associations originate in molecular clouds that are restructured through feedback from the stars, such as winds and explosive mass ejections  \citep[][]{Lada2003, Colin2013, Kruijssen2019a, Adamo2020, Kim2021}.  Whatever the history of this interaction is, once free from the immediate natal gas, a young cluster still finds itself embedded in the larger scale, structurally complex environment from which the parent cloud emerged. This environment is dynamically chaotic since turbulence produces density and velocity fluctuations on a wide range of length scales \citep[e.g.,][]{Vazquez-Semadeni1994, Federrath2013, Chevance2020}. These dense, evanescent turbulent structures generate a fluctuating potential that will harass the cluster and alter its evolution from that expected if it were isolated \citep{Kruijssen2011a}. Before exiting from the formation region, the cluster has to pass through this denser environment. Here it will experience a variable potential that can distort its spatial distribution and increase its global velocity dispersion \citep[see, e.g.,][]{Gieles2006}.

In this work, we study how this gaseous environment affects a cluster's dynamical evolution through numerical models that treat the hydrodynamics of the gas and the dynamics of the stars simultaneously. 
We focused on young massive clusters (YMC), that display ideal characteristics to analyse the effect. The majority of them originates from massive star-forming complexes \citep{Grosbol2012, Kruijssen2019a}, where the high densities and the structures created by supersonic motions are likely to have a more prominent impact on their evolution.
In addition, their short evolutionary timescale allows us to study the presence of coupling between the processes.

\section{Simulation methods}
\label{sec:methods}

To simulate the interaction between a star cluster and its surrounding environment, each component requires a different approach ($N$-body and hydrodynamical, respectively). We used specialised codes to follow their evolution and handled the interaction between them with \texttt{AMUSE} \footnote{Astrophysical Multipurpose Software Environment} \citep{PortegiesZwart2009,Pelupessy2013,PortegiesZwart2018}. We adopted the $N$-body code \texttt{PeTar} \citep{Wang2020}, which combines algorithmic regularisations, particle--particle and particle--tree techniques \citep{Kustaanheimo1965,Ahmad1973,Barnes1986}, for the cluster dynamics, and the smoothed-particle hydrodynamics (SPH) codes \texttt{GADGET-2} \citep{Springel2005} and \texttt{Fi} \citep{Gerritsen1997,Pelupessy2004} to treat the environment.
We performed four different sets of simulations: clusters with equal mass stars, clusters with an initial mass function (IMF), and each of these with and without the environment. For convenience, we marked pure $N$-body reference runs (i.e, without environment) with the suffix \texttt{Nb}. We describe the choice of the initial conditions below.

\subsection{Star cluster}
\label{sec: initial_param_clust}
We chose the physical parameters to resemble a typical YMC \citep[see, e.g.,][]{PortegiesZwart2010}, with the total mass of each model $M = 10^4\,\MSun$.
To study the effects of the interplay of the dynamical processes in the cluster and the environment, we chose five different IMFs: \citet{Salpeter1955} with $0.15 \leq m/\MSun \leq 3$ and with $0.15 \leq m/\MSun \leq 6$; \citet{Kroupa2001} with $0.1 \leq m/\MSun \leq 3$ and with $0.1 \leq m/\MSun \leq 6$ (the change from $0.15$ to $0.1  \,\MSun$ yields a more similar mean mass to the models with the Salpeter IMF); and an equal-mass stellar population with $m = 0.38\,\MSun$, which is the mean mass of the narrower Salpeter distribution. For the star cluster sizes, we adopted three different values of the virial radius, $\rvir = 0.7$, $1.3$ and $3.0\,\pc$. We generated the stellar positions and velocities according to the \citet{Plummer1911} profile.
The corresponding core density, $\rhoc$, and all combinations of the initial conditions used are listed in Table~\ref{tab:Nbody_simulations}. The models names are a combination of $\rvir$ and the IMF with the upper mass limit (the name contains the suffix ``\texttt{\_Nb}'' when the model evolves only as an isolated $N$-body system).

The observed relative motions between clusters in star-forming regions are of the order of a few $\kms$, leading to an evolution inside the dense region of a few tens of Myr \citep[see Sec.~\ref{sec: initial_param_hydro};][]{Elson1987,Kuhn2019,Roccatagliata2018,Roccatagliata2020}. Therefore, we followed the evolution of each star cluster for ${\gtrsim}100\,\Myr$, which would be a conservative choice to let the system escape the high-density environment. We did not include stellar evolution since even the most massive stars in our models (i.e., $6\,\MSun$) would have the main sequence phase of a similar length as the time span of our simulations \citep[cf.][]{Kippenhahn2012}.

\subsection{Environment}
\label{sec: initial_param_hydro}

We modelled the region surrounding the star cluster with \texttt{GADGET-2} using equal-mass SPH particles, initially distributed uniformly and at rest in a box of size $L=400\, \pc$, with periodic boundary conditions. The mean number density and the mean molecular weight were the same in all models, that is, $\langle n\rangle = 10\,\ccm$ and $\mu = 1.27$, respectively. These values are typical for giant molecular clouds complexes \citep[see, e.g.,][]{Blitz1980,Elmegreen2007,Heiner2008,Heiner2008a,Ballesteros-Paredes2020}. With the resolution adopted in the reference simulation ($128^3$ SPH particles), these parameters lead to a mass for each gas particles of $9.3\,\MSun$. In addition, we introduced the gas thermal behaviour according to the method described in \citet{Vazquez-Semadeni2007}, adopting the parametric heating and cooling functions from \citet{Koyama2002}.

We gradually developed a divergence-free turbulent field by injecting a fixed amount of energy at each integration time-step, giving a random velocity kick to each gas particle. We defined this energy input using a prescription from \citet{MacLow1999a} such that the equilibrium with the turbulent dissipation would be reached when the box had a velocity dispersion of $\Vrms$ (see Table~\ref{tab:SPH_simulations} for the exact values in the simulations). We generate the kicks according to a Gaussian random field having flat spectrum only in between $\kkick \leq k \leq \kkick+1$, where we randomly chose the value of $\kkick$ at each time-step from a uniform distribution in the range listed in Table~\ref{tab:SPH_simulations}.

We did not include the self-gravity of the gas. However, the supersonic turbulent motions (with $\Vrms\approx15$ times the sonic speed at equilibrium for gas densities of $n=10\,\ccm$) and the thermal instability \citep{Field1965} allowed by the heating and cooling functions are sufficient to create dense structures in the gas. Moreover, in our set-up, we do not expect gravity to play a crucial role on the large scale structures (see also, e.g., \citealt{Klessen2000} for an in-depth study of the relative importance of gravity and supersonic motions in highly turbulent environments.).
Further details of the environmental setup are in \citet{suin_thesis}.

\subsection{Cluster and environment}

In the environmental simulations of clusters, we first let the environment fully develop the turbulent field. Once the velocity dispersion of the box stabilises on the desired $\Vrms$, we insert the star cluster.
To avoid the phase space of the simulations becoming too wide, we limited this part of the study to a single set of parameters of the background ($\Vrms=15\,\kms$ and $\kkick\in[4,8]$, using a resolution of $128^3$ SPH particles; see Table~\ref{tab:SPH_simulations}). This setup mimics the mean velocity dispersion \citep{MacLow2004, Dib2006,  Klessen2010,  Kritsuk2017} and driving wavelength of the physical regions we are schematising \citep[cf.][]{Norman1996, Scalo2004, Brunt2009}. We also check that there are no massive structures nearby as they could alter the dynamical evolution significantly -- we required that the mass of gas included within two virial radii was $\lesssim 5\,\%$ of the cluster mass.
The systems then continue to evolve together, with the cluster perceiving the potential generated by the environment. We treated this interaction using the \texttt{Bridge} routine implemented in \texttt{AMUSE}, and letting the code \texttt{Fi} \citep{Gerritsen1997, Pelupessy2004} generate the potential field of the gas \citep{Rieder2022}. We note that the domain of the gaseous environment is periodic. Thus, stars escaping the cluster at higher velocities can depart arbitrarily far from the cluster and still feel the generated potential.

\section{Results}
\label{sec:results}
We describe the impact of the surrounding environment on the star cluster evolution in three separate ways. First, we focus on the effect of environmental harassment on the core evolution. In the second part, we analyse the cluster outer density distribution. Finally, we look at the overall stellar mass loss.

\subsection{Core evolution}
\label{sec:core_evolution}
\begin{figure*}
    \centering
    \includegraphics[width=0.49\linewidth]{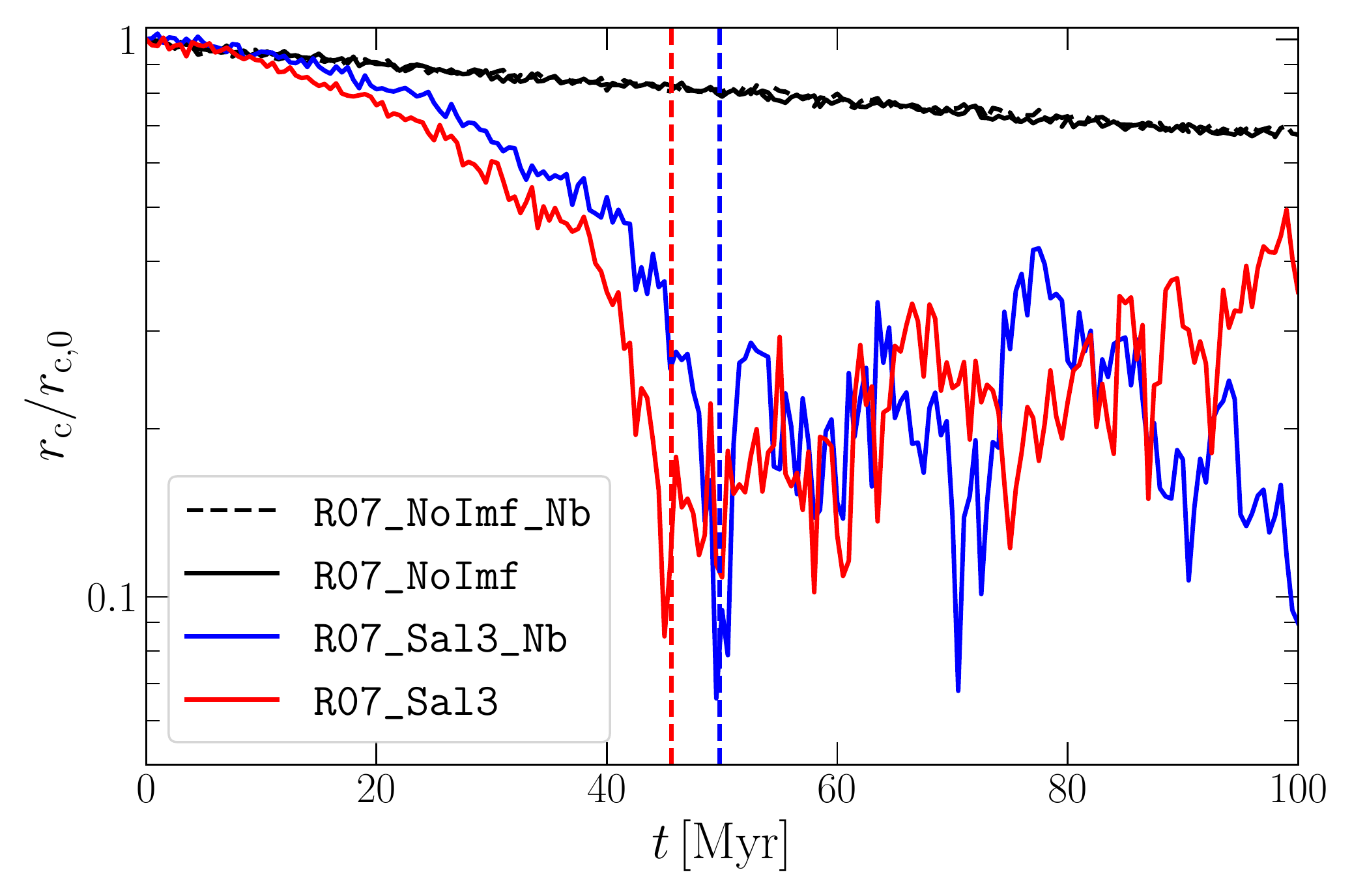}
    \hfill
    \includegraphics[width=0.49\linewidth]{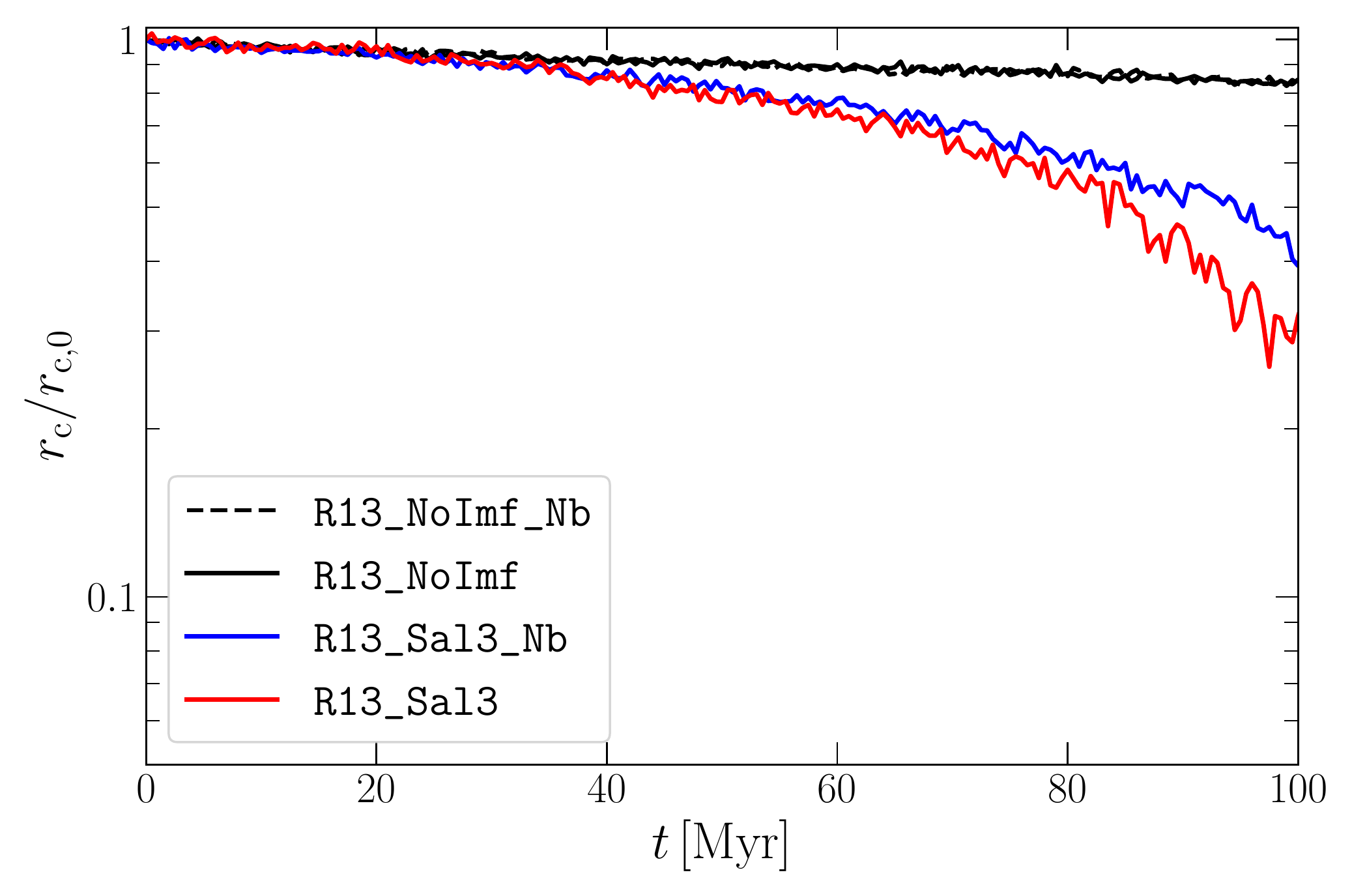}
    
    \caption{Evolution of the core radius, see Eq.~\eqref{eq: core_radius_definition}, normalised to the initial core radius. The plots show the clusters with the Salpeter IMF or equal-mass stars, both isolated and embedded into the environment (as labelled in the legends, we note that the two black lines are almost superimposed). The left-hand plot shows the more compact models with $\rvir=0.7\,\pc$, the right-hand one is for $\rvir=1.3\,\pc$. The moments of core collapse are highlighted with a vertical line.
    }
    \label{fig: core_radius_evolution}
\end{figure*}
Although the tidal force acting on the stars increases linearly with the distance from the cluster centre, the environment also has a profound effect on the inner parts of the cluster -- its presence leads to a quantitative acceleration of the core contraction. Tidal shocks increase the velocity dispersion of the cluster so that more stars evaporate from the core to the outer halo. This acts in the same direction as the dynamical relaxation, and the evolution of the core speeds up \citep{Spitzer1987}. We can see this for selected models in Fig.~\ref{fig: core_radius_evolution} which shows the evolution of the core radius, defined as 
    \begin{equation}
        \rcore = \sqrt{ \frac{ \sum_{i}{\rho^2_i r^2_i} }{ \sum_{i}{\rho^2_i} } }
        \label{eq: core_radius_definition}
    \end{equation}
\citep[cf.][]{Aarseth2003}. Here $\rho_i$ is the density estimator near the \hbox{$i$-th} cluster member, computed with the \texttt{Hop} clump-find algorithm \citep{Eisenstein1998} using its nearest 12 neighbours, and $r_i$ is its distance from the cluster density centre \citep{Casertano1985}. Both compact models with an IMF experience core collapse (highlighted with a vertical line in the left-hand plots of Fig.~\ref{fig: core_radius_evolution}), but the environmental run reaches it sooner. This faster contraction of the core is also evident in the models where core collapse did not happen within the simulated temporal interval (see how the lines spread in the right-hand plot). Moreover, the plots show that the presence of a background has almost no effect on the core evolution of the equal mass clusters (the black lines remains superimposed during the whole simulation). The same is true during the very first phases of evolution in cluster with IMF (up to $10\,\Myr$ in the compact cluster and $30\,\Myr$ in the other). 

Although useful for tracking the temporal development of the core, Eq.~\eqref{eq: core_radius_definition} is a functional form that differs from the actual core radius in a way that depends on the shape of the cluster itself (see, e.g., Table~2 in \citealt{Casertano1985}). To identify the moment of core collapse, $\tcc$, we followed the self-similar argument of \citet{Lynden-Bell1980}. The core radius evolves as
    \begin{equation}
        \rcore(t) \propto \left( \tcc-t \right)^{2/(6-\alpha)} \,,
        \label{eq: core_radius_evolution_to_fit}
    \end{equation}
for $t\leq\tcc$, where $\alpha$ is a fitting parameter determined through numerical simulations. To apply this equation, we used the procedure of \citet{Pavlik2018} who show that before core collapse, the minima of Lagrangian radii are a fixed multiple of $\rcore$. Consequently, the evolution of these minima follows Eq.~\eqref{eq: core_radius_evolution_to_fit}\footnote{During their evolution, Lagrangian radii exhibit large fluctuations. These are particularly important in the inner regions, where the number of star is small, leading to a high statistical noise. To achieve a greater accuracy in the localisation of the minima, we smoothed the Lagrangian radii, using \citet{Savitzky1964} smoothing algorithm with a second degree polynomial, as in \citet{Pavlik2018}. The smoothing window corresponded to 9 data-points, i.e., $4.5\, \Myr$. The algorithm is implemented in \texttt{Python} as \texttt{scipy.signal.savgol\_filter}.}\!.%

Tab.~\ref{tab:fit core collapse} displays the results of the fit for the core collapsed clusters. In all simulations, we recover the expected value of $\alpha\approx2.21$ \citep[e.g.,][]{Takahashi1995, Lynden-Bell1980} within $3\,\sigma$ uncertainties. However, there is a tendency towards lower values in the simulations with an upper mass limit of $6\,\MSun$. This is compatible with the result of \citet{Pavlik2018}, who found $\alpha\approx1.5$ using a Salpeter IMF with an about ten times wider range than ours. The \texttt{R13\_Sal6\_Nb} model shows the greatest error in both fit parameters. In this simulation, the core contraction is less pronounced, which makes the location of the Lagrangian radii minima extremely sensitive to random fluctuations. Concerning the moment of collapse, Tab.~\ref{tab:fit core collapse} shows that the gaseous background accelerates the cluster evolution. Environmental simulations display smaller $\tcc$ with respect to their $N$-body counterpart. The only exception is run  \texttt{R07\_Sal6}, for which the dynamical timescale is short and the effect of the environment is negligible. 

We also analysed the binary ejected by the clusters after core collapse. There is a good correlation between the formation of the first tight binary \citep[cf.][]{Fujii2014,Pavlik2018}. We repeated the simulation \texttt{R07\_Sal6} four times with the environment and eight without increasing the size of the ejected binaries sample. We did not detect any statistically significant difference between the binding energy distributions extracted from the two cases. We refer to \citet{suin_thesis} for the in-depth study of the binary characteristics.

\begin{table}
	\centering
	\caption{List of fit results. The first column identifies the run (with \texttt{Nb} highlighting the reference $N$-body runs), the second the time at which the core collapsed according to the fit algorithm and the third the exponent $\alpha$. We note that the higher uncertainties in run \texttt{R13\_Sal6\_Nb} are physical, see the text.}
	\begin{tabular}{lcc} 
		\hline\hline
		Name & $\tcc / \Myr$ & $\alpha$  \\
        \hline
        \texttt{R07\_Sal3}       & $45.6\pm0.2$    &  $2.29\pm0.23$
        \\
        \texttt{R07\_Sal3\_Nb}   & $49.6\pm0.3$    &  $2.32\pm0.17$
        \\
        \texttt{R07\_Sal6}       & $22.0\pm0.1$    &  $1.99\pm0.13$
        \\ 
        \texttt{R07\_Sal6\_Nb}   & $20.3\pm0.3$    &  $1.87\pm0.28$
        \\
        \texttt{R13\_Sal6}       & $53.1\pm0.4$    &  $2.00\pm0.27$
        \\
        \texttt{R13\_Sal6\_Nb}   & $58.2\pm3.0$    &  $2.32\pm0.41$
        \\
        \texttt{R13\_Kr6}         & $61.1\pm0.1$    &  $1.56\pm0.22$
        \\
        \texttt{R13\_Kr6\_Nb}     & $66.0\pm0.7$    &  $2.13\pm0.31$
        \\
		\hline
	\end{tabular}
	\label{tab:fit core collapse}
\end{table}

\subsection{Outer region}
\label{sec:outer_region}
\begin{figure}
    \centering
     \includegraphics[width=1\linewidth, trim={2cm 0.1cm 2cm 0},clip]{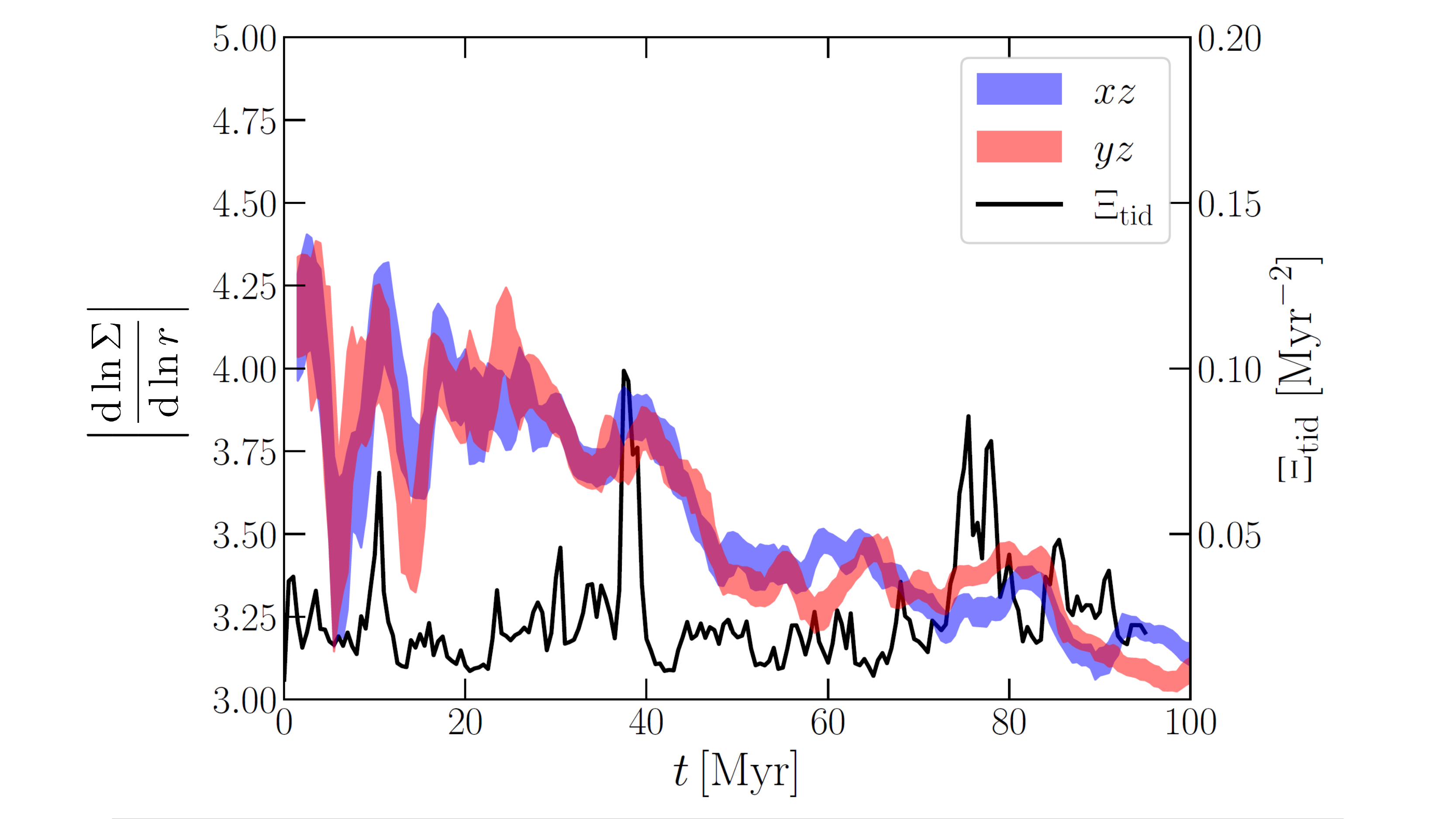}
    \caption{Temporal evolution of the modulus of the surface density exponent in the simulation \texttt{R07\_Sal3}, projected onto the planes $xz$ (blue) and $yz$ (red). We smoothed the data with a simple moving weighted average between 5 points ($2.5\,\Myr$). The black line shows the value of $\Xi_\tid$ (axes on the right), computed at the centre of the cluster. }
    \label{fig: beta_proj}
\end{figure}

We evaluated the impact of the background on the evolution of the radial density profile of the outer region. We chose the radial shells between the $94\,\%$ and $98\,\%$ Lagrangian radii with binning by $0.5\,\%$ and performed a least-square fit for the power-law 
            \begin{equation}
                \rho(r)\propto r^{-\beta}\,.
            \end{equation}  
We selected the lower bound to achieve an error smaller than $5\,\%$ for a Plummer model fit\footnote{For a Plummer model of scale radius $a$, $\beta=5 x^2/(1+x^2)$, where $x\equiv r/a$. Taking $r$ to be the $94\,\%$ Lagrangian radius, the exponent differs from its asymptotic value by $\approx0.045$}\!. The upper bound reduces the impact of statistical relative fluctuations  (bigger in the more sparse regions) on the fit.
The procedure implicitly assumes spherical symmetry. This, however, is justified by Fig.~\ref{fig: beta_proj}, which shows the behaviour of the surface density in run \texttt{R07\_Sal3} projected on the planes $xz$ and $yz$. The two sequences closely follow each other, so the three-dimensional nature of the problem only marginally affects the exponent $\beta$.

To compare the behaviour of $\beta$ with the tidal field acting on the cluster, we defined the variable $\Xitid(t)$ as the sum in quadrature of tidal tensor elements, computed at the cluster core centre~$\xcore$, 
    \begin{equation}
        \Xitid(t) \equiv \sqrt{\Xi_{\alpha\beta}(\xcore, t) \Xi_{\alpha\beta}(\xcore, t)}\,,
    \end{equation}
where 
    \begin{equation}
       \Xi_{\alpha\beta} (\xcore, t) \equiv  -\frac{\partial^2\phi}{\partial x_\alpha \partial x_\beta}(\xcore, t)\,,
        \label{eq: tidal force at distance drbeta}
    \end{equation}
and $\phi(\boldsymbol{x}, t)$ is the gravitational potential per unit mass generated by the gas at time $t$. We adopt the Einstein convention, summing over repeated indices. $\Xitid(t)$ has dimension of $\Myr^{-2}$ and is invariant under the change of coordinates, so it provides a good indicator of the strength of the tidal field affecting the cluster.
To give an idea of the physical value of $\Xitid$, a point-like object of mass $M_\mathrm{g}$ and distance $R_\mathrm{g}$ from the cluster would exert \hbox{$\Xitid=\sqrt{6}\, G M_\mathrm{g}/R_\mathrm{g}^{3}$}. In physical units, this scales as
    \begin{equation}
        \Xitid \sim 0.01\left(\frac{M_\mathrm{g}}{10^3\, \MSun}\right)\left(\frac{R_\mathrm{g}}{10\, \pc}\right)^{-3}\, \Myr^{-2}\,.
    \end{equation}
The trace, in this case, would be zero.

\begin{figure*}
    \centering
    \subfigure[\texttt{R07\_NoImf}]{\includegraphics[width=0.49\linewidth]{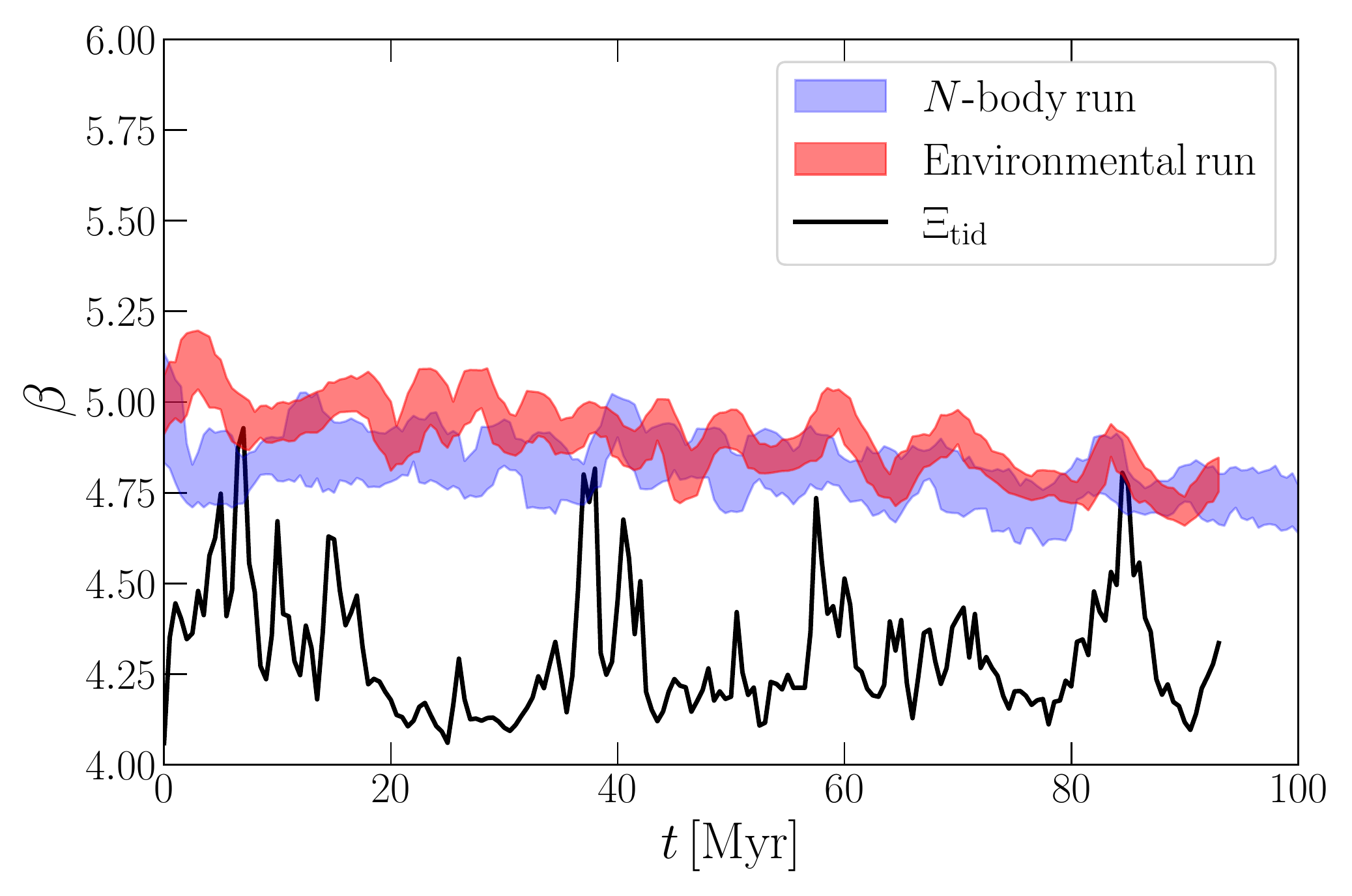} 
    \label{fig: beta_evolution - a}}
    \hfill
    \subfigure[\texttt{R07\_Sal3}]{\includegraphics[width=0.49\linewidth]{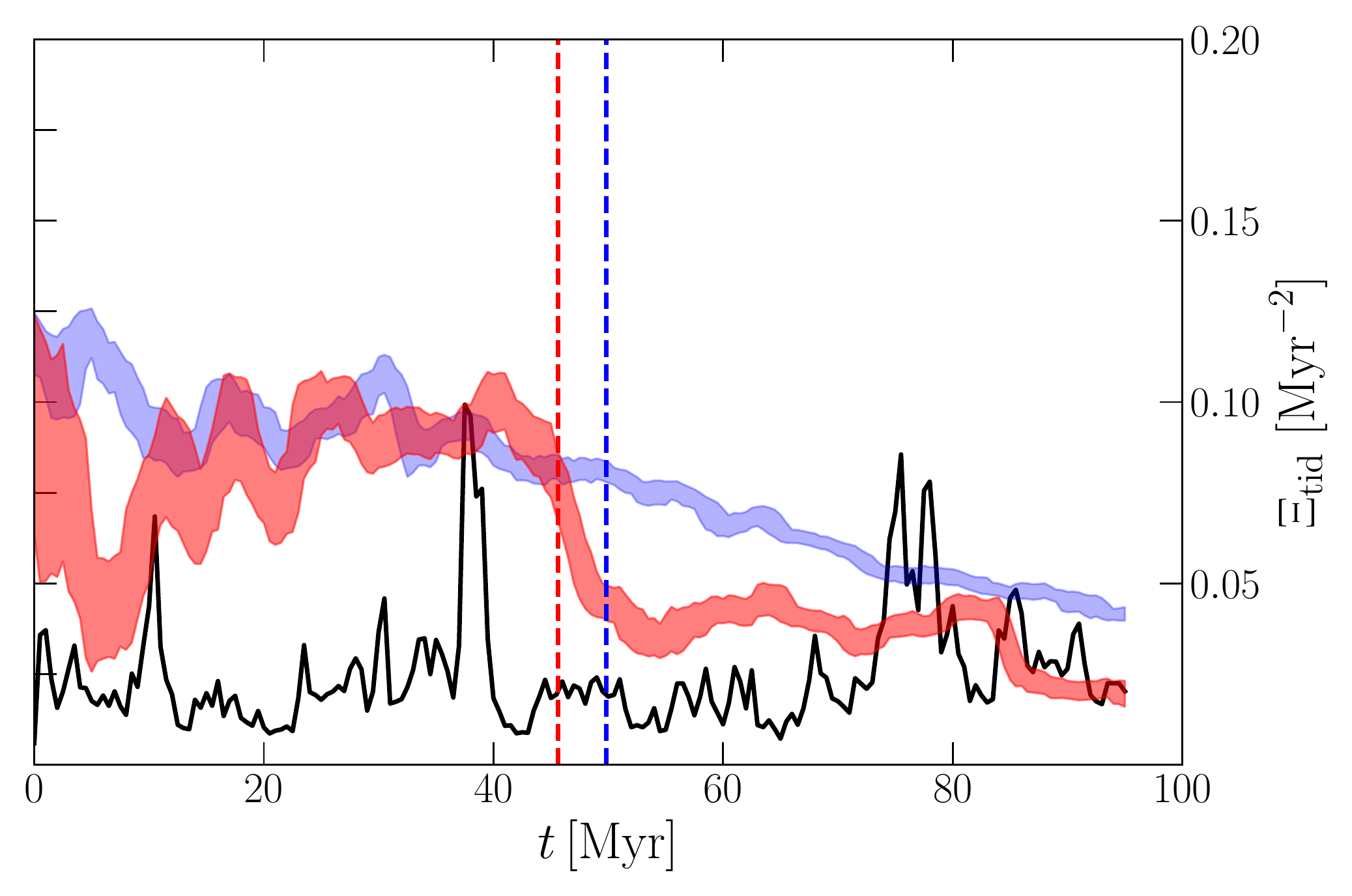}
    \label{fig: beta_evolution - b}}
    \\
    \subfigure[\texttt{R13\_NoImf}]{\includegraphics[width=0.49\linewidth]{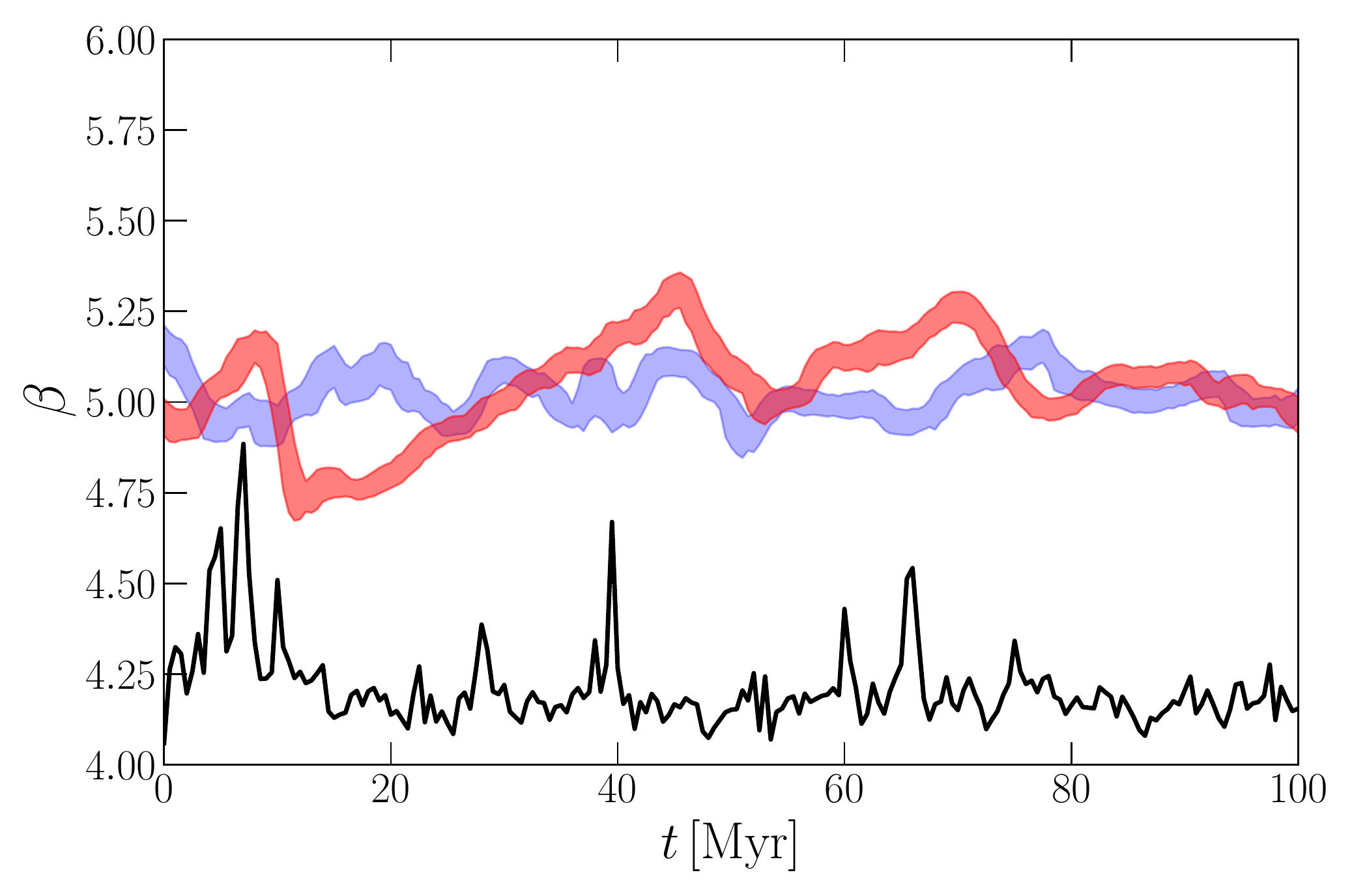}    
    \label{fig: beta_evolution - c}}
    \hfill
    \subfigure[\texttt{R13\_Sal3}]{\includegraphics[width=0.49\linewidth]{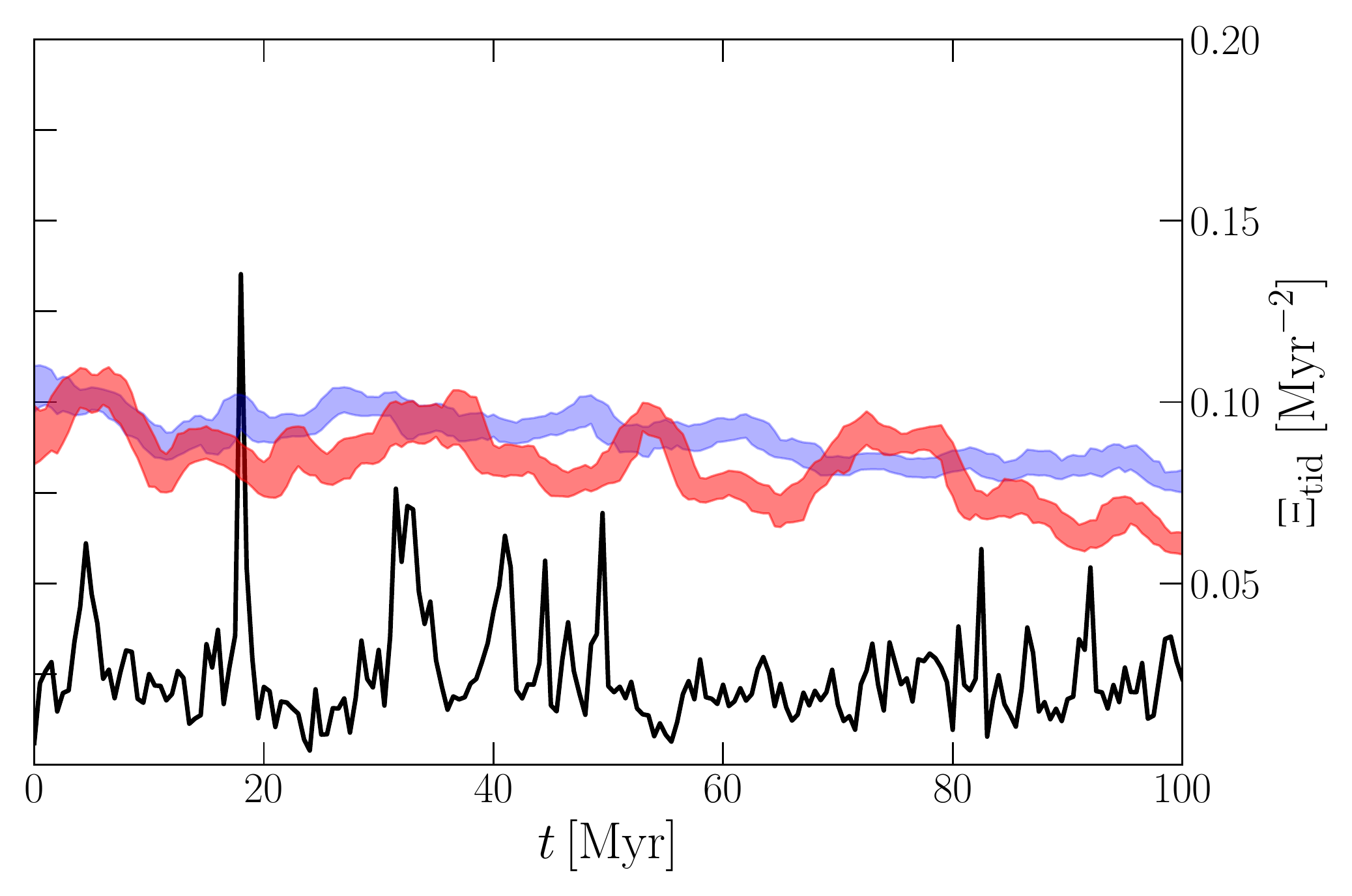}
    \label{fig: beta_evolution - d}}
    \\
    \subfigure[\texttt{R13\_Sal6}]{\includegraphics[width=0.49\linewidth]{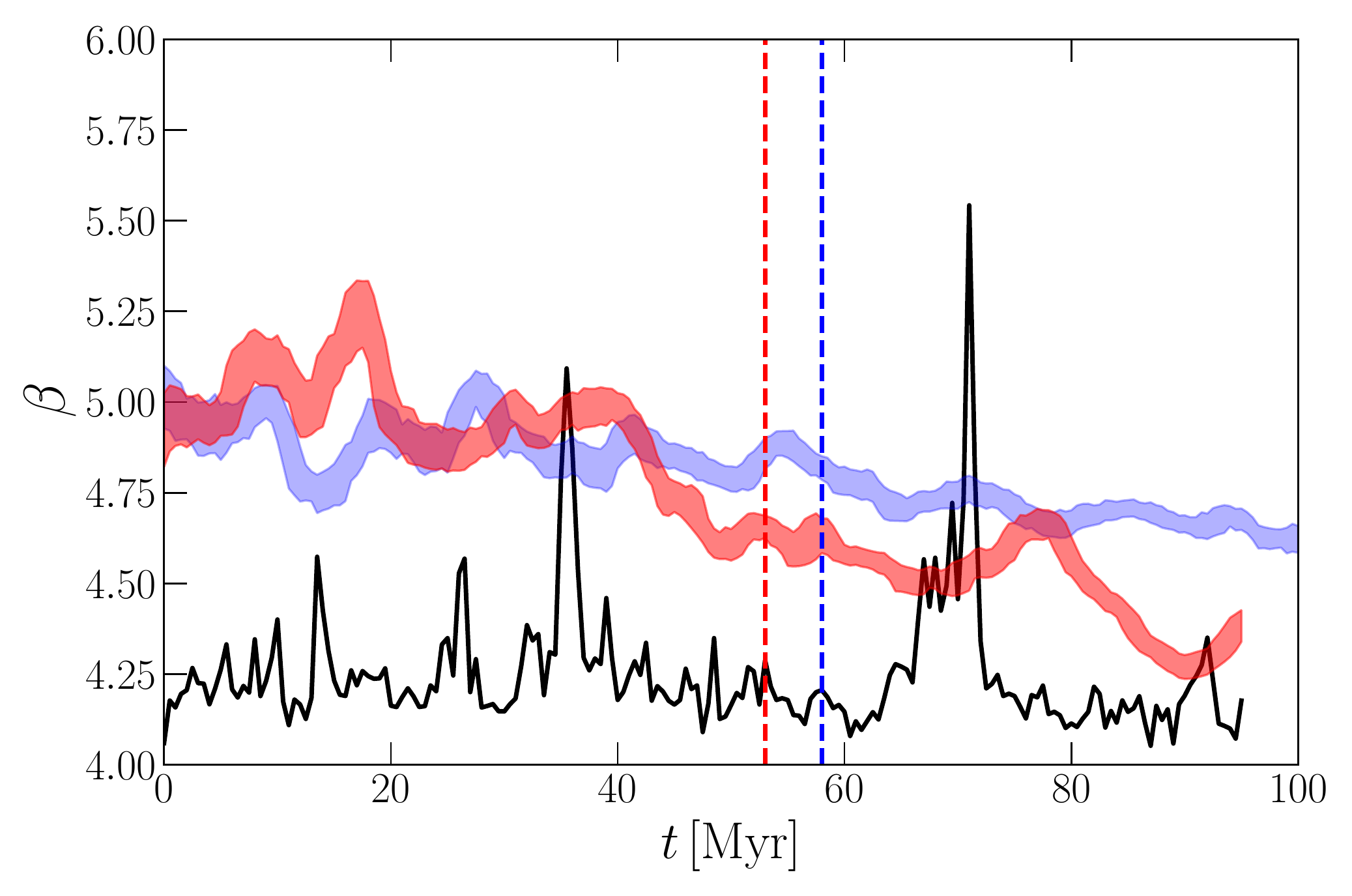}
    \label{fig: beta_evolution - e}}
    \hfill
    \subfigure[\texttt{R30\_Sal6}]{\includegraphics[width=0.49\linewidth]{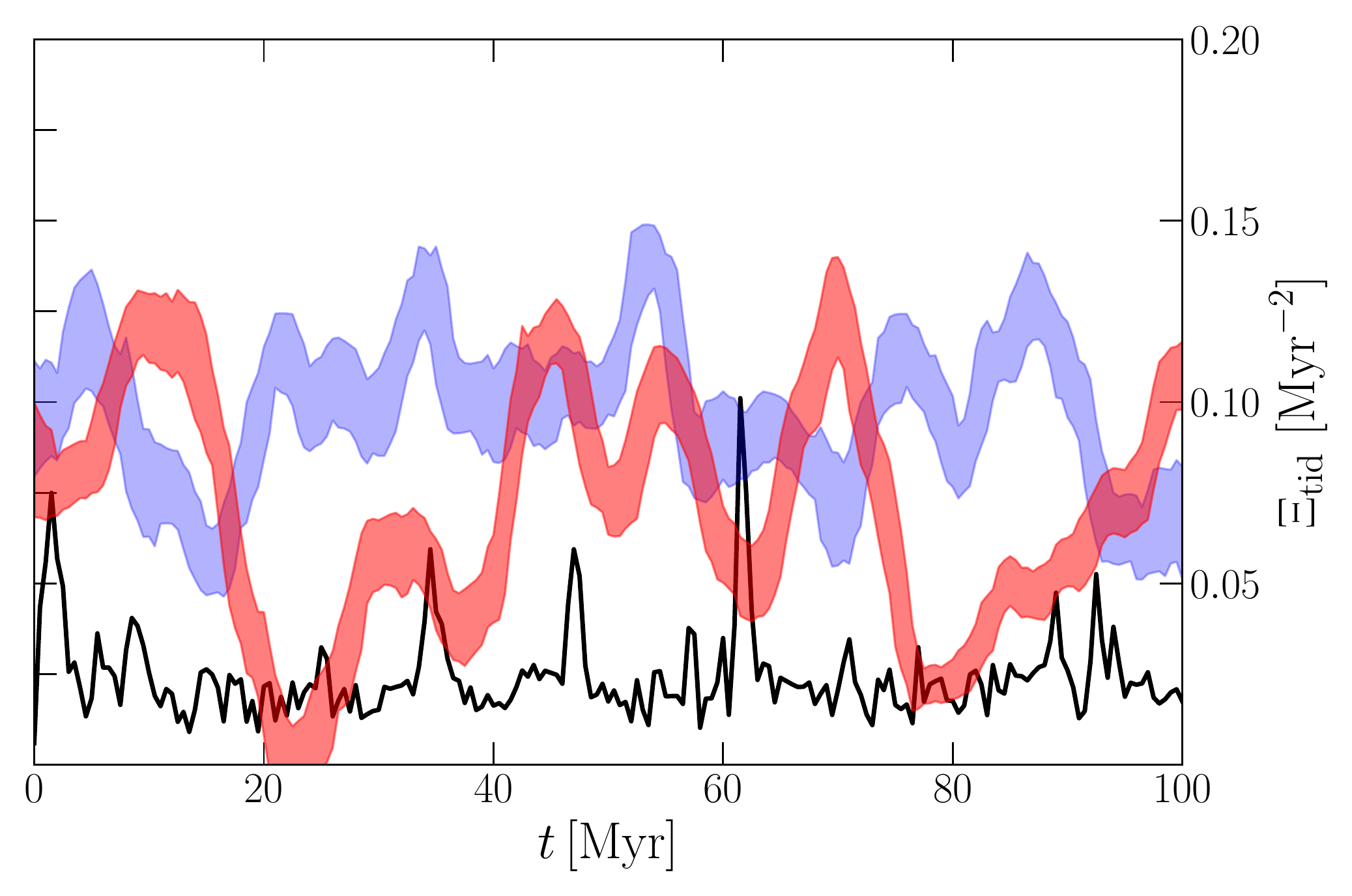}
    \label{fig: beta_evolution - f}}

    \caption{Evolution of $\beta$ throughout a selected set of simulations, with the axis on the left. The shaded areas identify the value of the fit within $1\sigma$ (blue for the $N$-body runs, red for those with environment). We smoothed the data with a simple moving weighted average of width $2.5\,\Myr$ (5~model outputs). The black line shows the value of $\Xi_\tid$ of the environmental runs, as in Fig.~\ref{fig: beta_proj} (axis on the right). The vertical dashed lines in the plots \subref{fig: beta_evolution - b} and \subref{fig: beta_evolution - e} highlight the moment of core collapse (see also Tab.~\ref{tab:fit core collapse}). }
    \label{fig: beta_evolution}
\end{figure*}

Fig.~\ref{fig: beta_evolution} shows the temporal evolution of $\beta$ in the simulations with and without the ambient gas, along with the indicator $\Xitid$ of the environmental run. In the isolated simulations, $\beta$ remains almost constant initially, fluctuating around the initial value of 5 (which comes from the Plummer model). It then starts to decrease approximately linearly in time while approaching core collapse.

Runs with the environment show greater fluctuations of $\beta$. In addition, we see that the effect of the background increases if the cluster is wider and its evolutionary phase more advanced. Large clusters are more sensitive to tides, and the core contraction sends stars out to the halo, where tides are stronger. A comparison between \texttt{R07\_NoImf} (Fig.~\ref{fig: beta_evolution - a}) and \texttt{R13\_NoImf} (Fig.~\ref{fig: beta_evolution - c}) allows us to discuss the first aspect. Without a mass function, the dynamical evolution of these clusters is almost negligible throughout the simulations (see the black lines in Fig.~\ref{fig: core_radius_evolution}), and we can extract the effect of the tides on the evolution directly. The clusters do not significantly differ from the evolution of their isolated counterpart. However, the fluctuations are bigger in both cases. The same can also be inferred from \texttt{R30\_Sal6} (Fig.~\ref{fig: beta_evolution - f}) -- in this case, the low stellar density slows down the mass segregation process and makes its dynamical development similarly slow as in the equal-mass case.
In addition, in \texttt{R13\_NoImf} (Fig.~\ref{fig: beta_evolution - c}) the trigger for the largest fluctuations in $\beta$ is the major encounter at $\approx10,\,40\,,65\,\Myr$, although in run \texttt{R07\_NoImf} it is not possible to extract a clear correspondence between close encounters (peaks in $\Xitid$) and variation of the power-law slope.  

Widening the mass range decreases the evolutionary timescale, and enhances the effect of the background. Core collapsed clusters (Fig.~\ref{fig: beta_evolution - a}-\ref{fig: beta_evolution - e}) display a more step-like evolution of the exponent $\beta$ when inserted into the environment. While approaching core collapse, the density gradient becomes more sensitive to the peaks in $\Xitid$. Close to these peaks, $\beta$ initially increases and the cluster becomes more compact. Then, stars populate the outer halo and $\beta$ decreases. The situation is similar in \citet{Gieles2006}, where the simplest example consists in a head-on encounter of the cluster with a spherical cloud. While approaching the cloud, the cluster contracts as stars suffer a net acceleration toward the centre along the direction perpendicular to the motion. After the encounter, the cluster expands due to the relative perpendicular velocities acquired by particles. The parallel velocity variation, instead, cancels out during the whole encounter. 

By the end of the simulation, the harassed clusters display softer outer power-law slopes. This also shows up in the last $\approx20\,\Myr$ of model \texttt{R13\_Sal3} (Fig.~\ref{fig: beta_evolution - d}), which at the end of the simulation is close to experiencing core collapse (as can also be deduced from Fig.~\ref{fig: core_radius_evolution}).
When far from core collapse, these clusters behave similarly to the equal mass case. At early stages of evolution, its shape did not change, and the environment does not affect the outer region permanently. Even though showing larger fluctuation than the reference runs, $\beta$ does not depart significantly from the initial value. As soon as a tidal shock occurs \textit{and} the cluster has an evolved configuration, $\beta$ departs from the linear behaviour of the isolated case (see the shocks at $\approx40,\,70\,\Myr$ and the subsequent decrease in $\beta$ in Fig.~\ref{fig: beta_evolution - a}--\ref{fig: beta_evolution - e}).

\subsection{Cluster mass loss}
To determine whether a star is bound to the cluster is not straightforward when an external potential is present. The cluster members are constantly interacting with the gaseous environment which can differentially accelerate them. While passing through or near a cloud, a star in the cluster gains enough kinetic energy to appear unbound, and the usual definition, which labels escaping stars based on their instantaneous binding energy to the system, fails. However, over a whole encounter, only the orthogonal velocity component remains high, while any velocity gains in the parallel component cancel out. To avoid the fluctuations in the mass loss due to these encounters, we only mark as ``escapers'' the stars that remain unbound throughout the last $5\,\Myr$ of the simulation (see Appendix~\ref{ap:mass_loss} for further justification of this choice).

\begin{figure}
    \centering
     \includegraphics[width=1\linewidth]{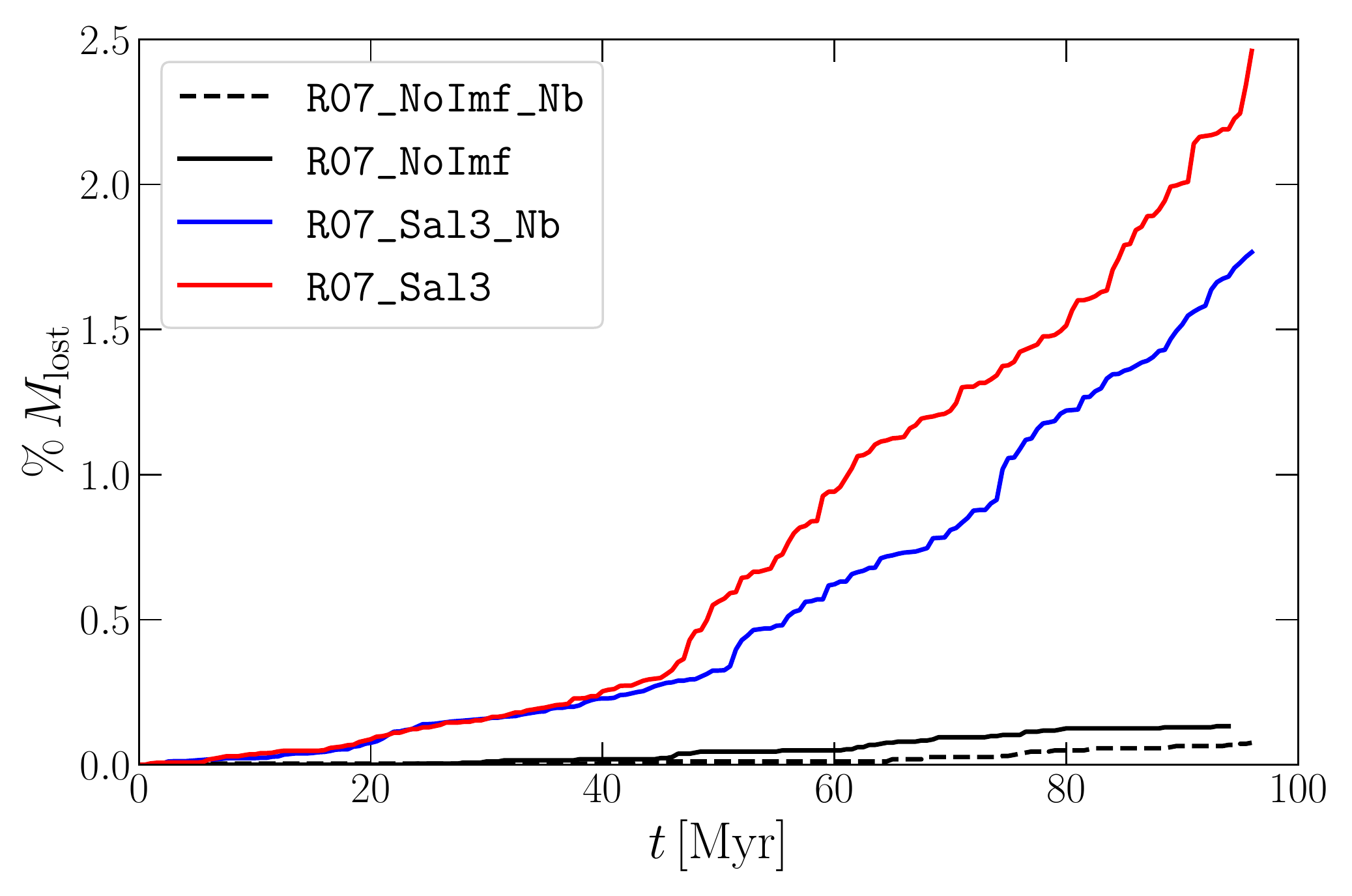}\\
     \includegraphics[width=1\linewidth]{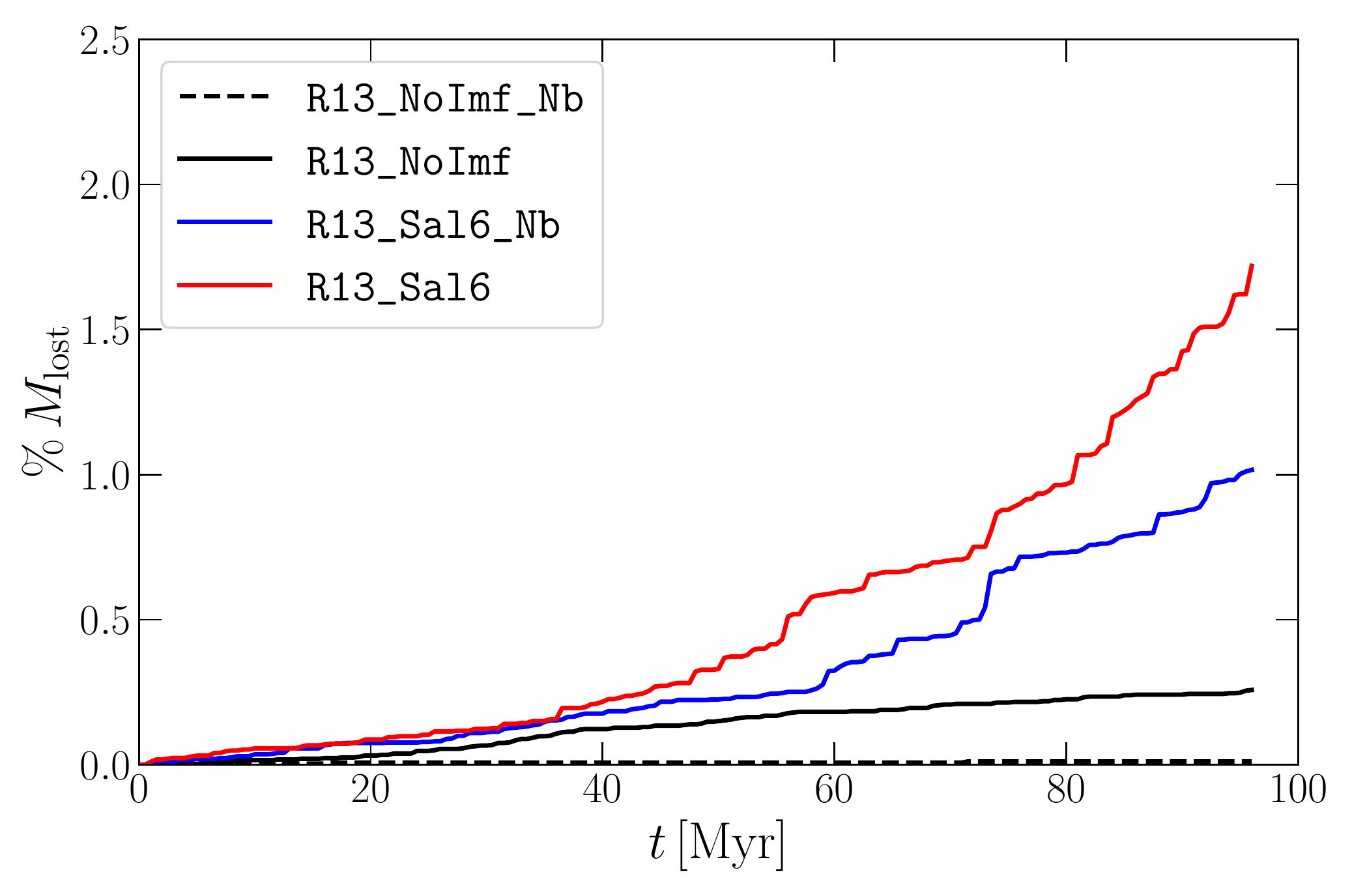}
     
    \caption{Evolution of the cluster mass loss in percents of the total initial mass (see the text for the definition).}
    \label{fig: mass_loss}
\end{figure}
Fig.~\ref{fig: mass_loss} shows two examples of the cumulative mass loss over time, comparing the simulations with and without the environment, along with the mass loss from the equal mass clusters. This highlights the presence of a coupling mechanism between the two evolutions. As for the outer density distribution, the coupling only shows up when both effects are present. The mass loss of run \texttt{R07\_Sal3} is particularly explanatory (see the top panel of Fig.~\ref{fig: mass_loss}), as it follows the isolated cluster until just before core collapse. In addition, after core collapse the mass loss presents a steeper increase than in the isolated cluster.

\begin{figure}
    \centering
    \includegraphics[width=1\linewidth]{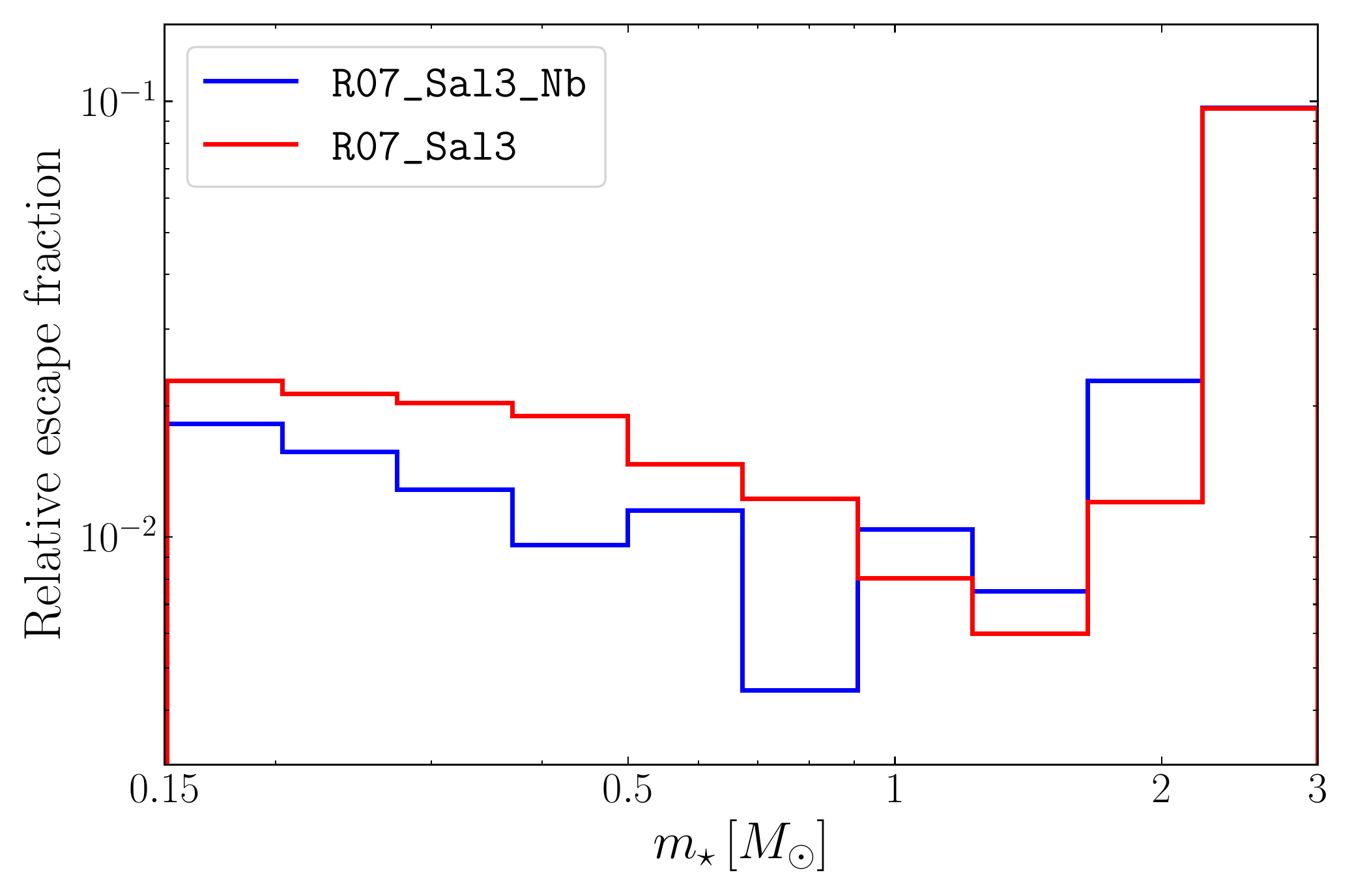}\\
    
    \caption{Mass distribution of escapers normalised to the initial IMF for the runs \texttt{R07\_Sal3} (red) and its $N$-body reference run (blue). }
    \label{fig:relative_loss}
\end{figure}

\begin{table}
  \centering
  \caption{Mass loss of simulated cluster computed at $95\,\Myr$. The first column is the name of the run with environment and IMF, in which mass loss is specified in the second column. That of the respective isolated cluster is shown in the second column, while the fourth and the fifth are the mass loss from simulation of the same cluster without IMF. }
  \renewcommand\arraystretch{1.05}
  \begin{tabular}{l c c c c}
    \hline\hline
    \multirow{2}{*}{Cluster} & \multicolumn{4}{c}{\%\,$M_\mathrm{lost}$}  \\
    \cline{2-5}
     & IMF &  IMF, \texttt{Nb} & No IMF &  No IMF, \texttt{Nb}  \\
		\hline
        \texttt{R07\_Sal3}        & $2.47$        &  $1.78$        & $0.14$   &    $0.08$
        \\
        \texttt{R07\_Sal6}        & $3.66$        &  $3.25$        & $0.14$   &    $0.08$
        \\
        \cline{4-5}
        \texttt{R13\_Sal3}        & $0.46$        &  $0.24$        &  $0.26$  &    $0.01$
        \\ 
        \texttt{R13\_Sal6}        & $1.73$        &  $1.02$        &  $0.26$  &    $0.01$
        \\
        \texttt{R13\_Kr3}          & $0.77$        &  $0.32$        &  $0.26$  &    $0.01$
        \\
        \texttt{R13\_Kr6}          & $1.47$        &  $1.22$        &  $0.26$  &    $0.01$
        \\
        \cline{4-5}
        \texttt{R30\_Sal3}        & $1.73$        &  $0.15$        &  $1.64$  &    $<0.01$
        \\
        \texttt{R30\_Sal6}        & $1.89$        &  $0.18$        &  $1.64$  &    $<0.01$
        \\
		\hline
  \end{tabular}
  \label{tab:mass_loss}
\end{table}

Tab.~\ref{tab:mass_loss} provides the mass lost in each simulation after $95\,\Myr$. In every simulation, the mass loss from the isolated equal mass cluster is negligible. This supports the choice of using the embedded equal mass cluster as a reference for the evolution caused only by environmental harassment. As expected, the importance of the tides increases as the cluster expands. In contrast, the mass lost in pure $N$-body simulations depends only on the mass segregation timescale, which is proportional to $\rho^{-1/2}$ and inversely proportional to the width of the IMF. 

Finally, it is worth noting that two-body relaxation and tidal interactions affect the IMF in different ways. The former causes mass segregation with higher-mass stars moving inward and lower-mass stars outward, which enables the lower-mass stars to evaporate from the cluster \citep[see also, e.g.,][]{Chandrasekhar1942, SpitzerLyman1969, Binney1987}. The cluster then forms a dense core of high-mass stars and once it starts to collapse, binaries form in the centre. These perturb their neighbours, and we begin to observe massive stars escaping from the mass segregated core \citep[see also][]{Hills1975, Hut1985a}.
On the other hand, if the relaxation timescales are long, the cluster does not segregate as quickly and the tidal harassment is the dominant reason for mass-loss. Stars are peeled-off from the outer regions independent of their masses.
Consequently, the relative contributions of these two processes determine the mass distribution of the escapers.  This is shown in Fig.~\ref{fig:relative_loss}, where the mass distribution of escaping stars divided by the original IMF in models \texttt{R07\_Sal13} and \texttt{R07\_Sal13\_Nb}. When the environment is present, the tidal shocks enhance fraction of escaping stars at intermediate masses. Instead, the loss of massive stars, similar in the two models, can be linked to the relaxation process ongoing in the core that is composed almost exclusively of massive stars and binaries at later times. In our model, the percentage of mass loss is small so this does not significantly affect the mass function of the remaining members. However, the same process happens in protoclusters still embedded in the parent cloud. Their gas density is orders of magnitude higher than our $n=10\,\ccm$, and structures are closer to the cluster \citep{Kruijssen2011a, Kruijssen2012}. At the same time, the IMF of massive protoclusters is much wider, since massive stars are still on the main sequence, so that the evolutionary timescale is also reduced \citep{Allison2009,Yu2011}. In the end, the effects we described could indeed play a crucial role when inferring the IMF of young clusters.

\section{Limitations}
\label{subsec:limitations}
The choice of initial conditions has trade-offs. Adopting the Plummer model and truncating the IMF at $6\,\MSun$ allows us to avoid the more complex stages of the cluster formation that a more physical model would require. However, stellar evolution and feedback play a significant role when including younger stars with higher masses. Massive stars sink faster toward the cluster centre and die earlier \citep{PortegiesZwart2007}. The presence of a denser environment also enhances tidal perturbations. Nevertheless, the mechanisms studied here will impact the evolution of the cluster.
Regarding the background setup, we mimicked the averaged observed values of star-forming regions. The constant energy input generated a constant turbulent field in which structures formed and dissolved. The feedback mechanisms acting in such environments are complicated and still far from completely understood. Adding self-gravity increases the computational times due to the collapsing high-density regions.
Our work is, instead, intended to show that even simple models have observational consequences and to provide insights regarding the interplay between the dynamical evolution of star clusters and the gaseous background
that will only be magnified by achieving more physical simulations.

\section{Conclusions}
\label{sec:conclusions}
We conducted a series of numerical simulations following the  evolution of both the star cluster and its surrounding ISM environment. Gaseous structures, created by supersonically driven turbulent motions, produced a spatially and temporally varying external potential which perturbed the embedded cluster. 

The core evolution and the external power-law slope of the density distribution both exhibited effects induced by the gas. The tidal shocks increase the fractions of core stars that are sent into the halo. This process acts similarly to dynamical relaxation. We note that this sped-up redistribution of energy can also continue to later stages of the cluster evolution, every time the cluster passes through a star-forming region. This supports the finding of \citet{Gnedin1999_disk_shock_GC} of a reduced core collapse timescale using spherical Fokker--Planck models.

For the cluster periphery, we found that the exponent of the asymptotic density distribution, $\beta$, displays greater fluctuations induced by the environment. While approaching core collapse, tides significantly affect $\beta$. The slope departs from the almost linear temporal evolution expected in the isolated case, displaying a more step-like behaviour. In addition, all core collapsed runs end up with a lower value of $\beta$ as a result of the accelerated evolution. Both effects would contribute to observing more clusters in a more relaxed and expanded configuration \citep[e.g.,][]{PortegiesZwart2010, Banerjee2017}.

This coupling also manifests in the cluster mass loss. In each environmental simulation, the final mass loss is larger than the sum of the respective reference runs. This highlights a coupling between the dynamical evolution and the presence of the environment.

\begin{acknowledgements}
We thank Maurizio Davini and the Direzione Infrastrutture Digitali of the University of Pisa for providing access to two 128-core DELL R630i servers of the High-Performance Computing Division of the San Piero a Grado Green Data Center, without which this work would not have been possible.
We are grateful to Veronica Roccatagliata and Michele Cignoni for  insightful discussions. We also thank the \texttt{AMUSE} team for the support provided.
This research has made use of NASA's Astrophysics Data System Bibliographic Services.
The \texttt{Python} programming language with \texttt{NumPy} \citep{numpy} and \texttt{Matplotlib} \citep{matplotlib} were used in this project.
\end{acknowledgements}

\bibliographystyle{aa}
\bibliography{bibliography}

\clearpage
\appendix
\onecolumn

\section{Model parameters}
\label{ap:params}
\setlength{\tabcolsep}{5pt}
\begin{table*}[!h]
\begin{minipage}{0.57\linewidth}
	\centering
	\caption{List of $N$-body simulations of the star cluster. For each one, the following initial parameters are given: the virial radius, the core density, the IMF and its mass range, and the mean stellar mass. We note that the total mass of each model is $10^4\,\MSun$.}
	\begin{tabular}{lccccc}
		\hline\hline
		Name & $\rvir/\pc$ & $\rhoc/(\MSun\, \pc^{-3})$ & IMF & $m/\MSun$ & $\langle m \rangle/\MSun$ \\
		\hline
		\texttt{R07\_NoImf} & $0.7$ & $3.5\cdot10^4$  & eq.~mass & ---        & $0.38$ \\
		\texttt{R07\_Sal3}  & $0.7$ & $3.5\cdot10^4$  & Salpeter & $[0.15,3]$ & $0.38$ \\
		\texttt{R07\_Sal6}  & $0.7$ & $3.5\cdot10^4$  & Salpeter & $[0.15,6]$ & $0.42$ \\
		\texttt{R07\_Kr6}   & $0.7$ & $3.5\cdot10^4$  & Kroupa   & $[0.1,6]$  & $0.48$ \\
		\texttt{R13\_NoImf} & $1.3$ & $5.5\cdot10^3$  & eq.~mass & ---        & $0.38$ \\
		\texttt{R13\_Sal3}  & $1.3$ & $5.5\cdot10^3$  & Salpeter & $[0.15,3]$ & $0.38$ \\
		\texttt{R13\_Sal6}  & $1.3$ & $5.5\cdot10^3$  & Salpeter & $[0.15,6]$ & $0.42$ \\
		\texttt{R13\_Kr3}   & $1.3$ & $5.5\cdot10^3$  & Kroupa   & $[0.1,3]$  & $0.43$ \\
		\texttt{R13\_Kr6}   & $1.3$ & $5.5\cdot10^3$  & Kroupa   & $[0.1,6]$  & $0.48$ \\
		\texttt{R30\_NoImf} & $3.0$ & $5.0\cdot10^2$  & eq.~mass & ---        & $0.38$ \\
		\texttt{R30\_Sal3}  & $3.0$ & $5.0\cdot10^2$  & Salpeter & $[0.15,3]$ & $0.38$ \\
		\texttt{R30\_Sal6}  & $3.0$ & $5.0\cdot10^2$  & Salpeter & $[0.15,6]$ & $0.42$ \\
		\hline
	\end{tabular}
	\vspace{22pt} 
	\label{tab:Nbody_simulations}
\end{minipage}
\hfill
\begin{minipage}{0.4\linewidth}
	\centering
	\caption{List of SPH simulations of the environment. For each one, the following initial conditions are given: the root mean square velocity, the range of the kick wavenumber, and the number of SPH particles. We note that the mean number density is $\langle n\rangle = 10\,\ccm$ and the simulation box size is $L=400\,\pc$ in all models.}
	\begin{tabular}{p{6em}ccr} 
		\hline\hline
		Name & $\Vrms/\kms$ & $\kkick$ & $\Nsph$ \\
		\hline
		\texttt{V10\_k[4-8]}                     & 10 & [4, 8] & $128^3$ \\
		\texttt{V15\_k[4-8]\_lr}\tablefootmark{a}& 15 & [4, 8] & $64^3$  \\
		\texttt{V15\_k[4-8]}\tablefootmark{b}    & 15 & [4, 8] & $128^3$ \\
		\texttt{V15\_k[4-8]\_hr}\tablefootmark{a}& 15 & [4, 8] & $256^3$ \\
		\texttt{V15\_k[2-4]}                     & 15 & [2, 4] & $128^3$ \\
		\texttt{V15\_k[8-16]}                    & 15 & [8,16] & $128^3$ \\
		\texttt{V20\_k[4-8]}                     & 20 & [4, 8] & $128^3$ \\
		\texttt{V40\_k[4-8]}                     & 40 & [4, 8] & $128^3$ \\
		\hline
	\end{tabular}
	\tablefoot{
		\tablefoottext{a}{Labels ``\texttt{lr}'' and  ``\texttt{hr}'' identify the low and high resolution runs, respectively.}\\
		\tablefoottext{b}{\texttt{V15\_k[4-8]} is the model used in the simulations of star clusters embedded in the environment.}
	}
	\label{tab:SPH_simulations}
\end{minipage}
\end{table*}
\setlength{\tabcolsep}{6pt} 

\section{Stellar mass loss}
\label{ap:mass_loss}
\begin{figure}[h!]
    \centering
    \includegraphics[width =0.49\linewidth]{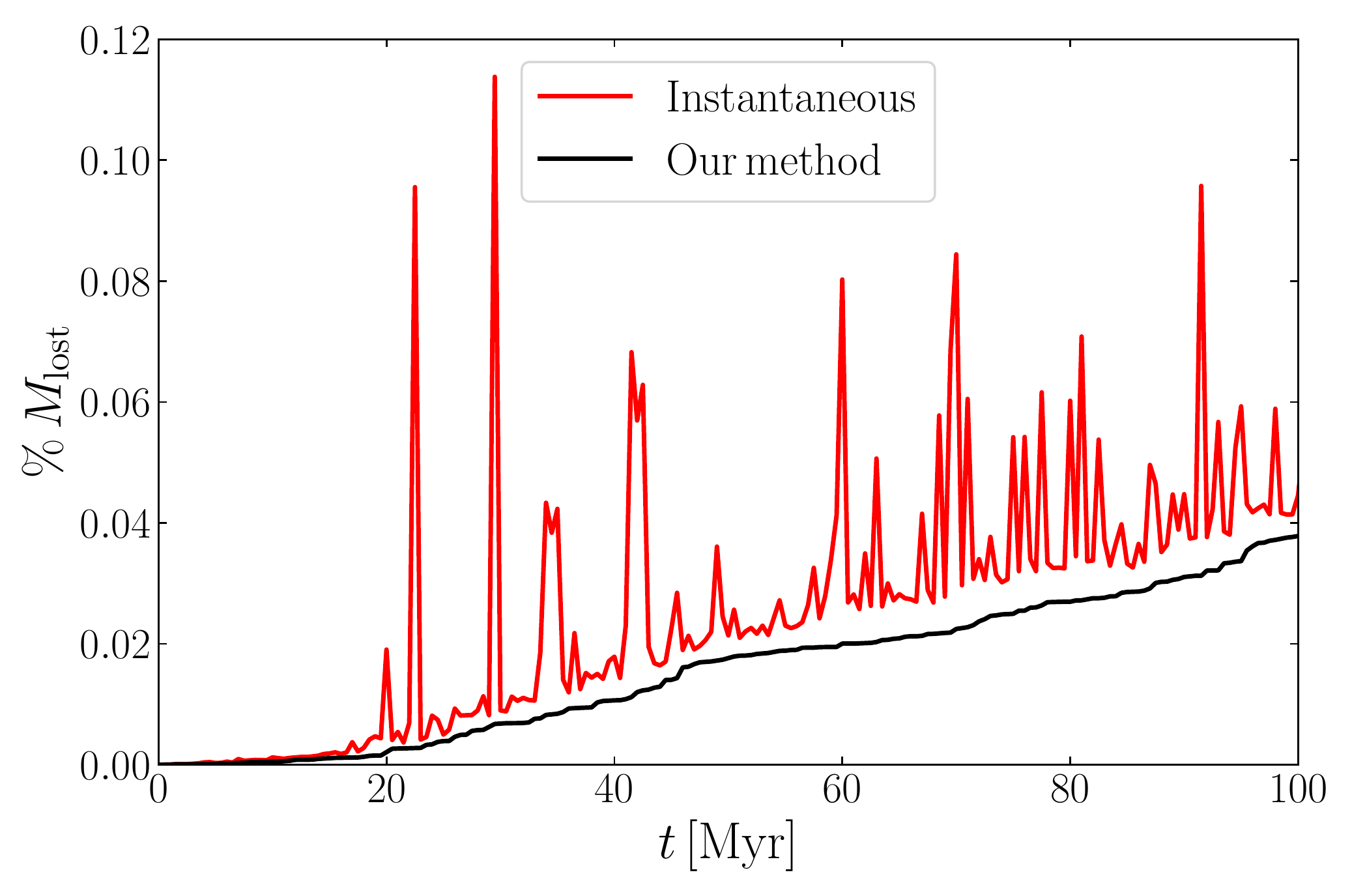}
    \hfill
    \includegraphics[width =0.49\linewidth]{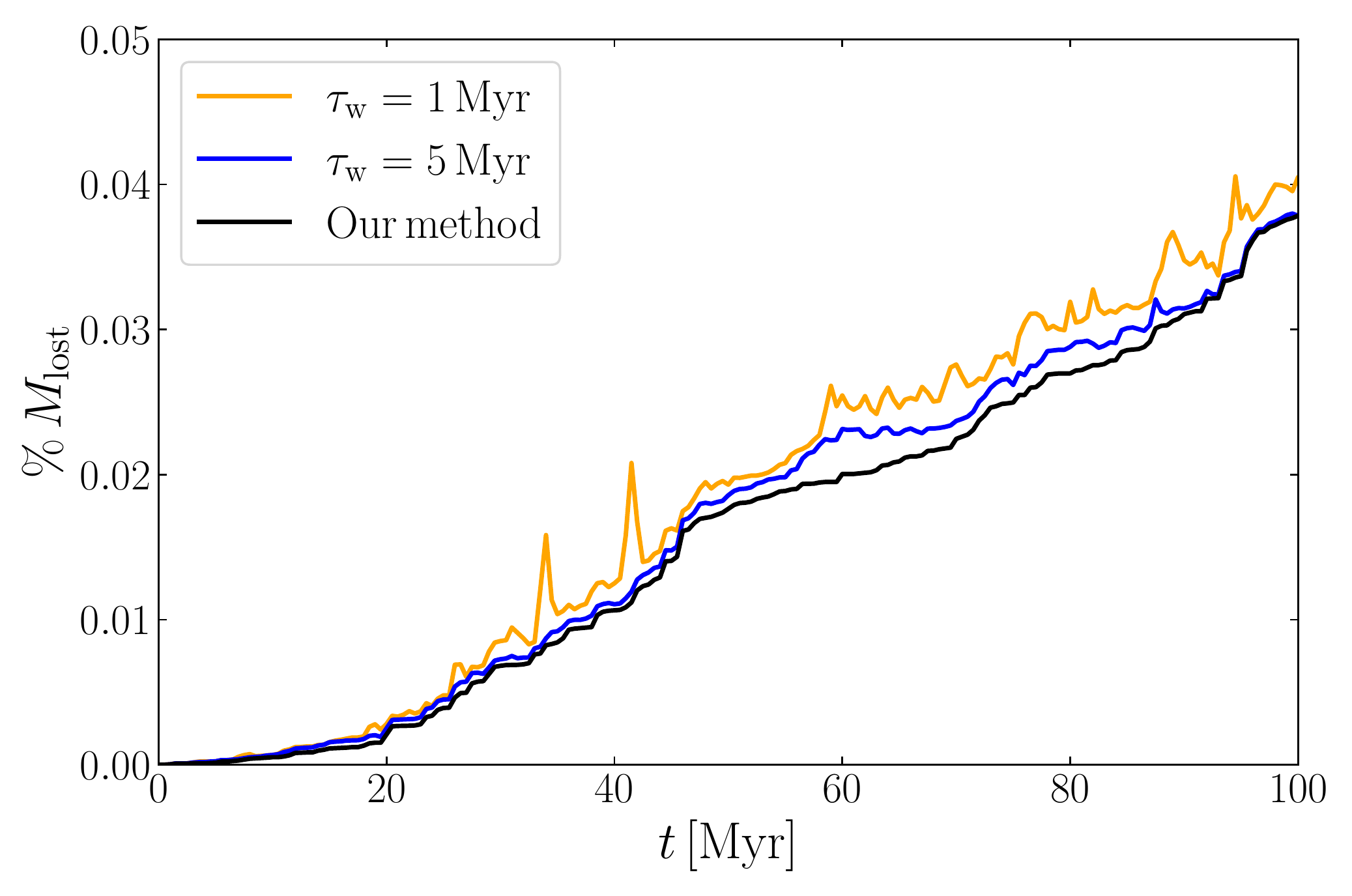}
    
    \caption{Comparison of the  methods to compute the fractional mass loss in model \texttt{R07\_Sal6}. \textbf{Left:} Comparison between our model (black) and the mass loss calculated instantaneously (red). \textbf{Right:} The alternative method proposed in the text compared to our model (black line), coloured lines correspond to different window lengths $\tauw$.}
   \label{fig: comparison methods unbound}
\end{figure}

Encounters with the environment increase the velocities of stars relative to the cluster centre, making them appear unbound for a short period of time (see the red line in Fig.~\ref{fig: comparison methods unbound}. The spikes, which show the instantaneous increase of the number of stars with positive energy, correspond to the moments when the cluster passes near a cloud. After the passage, most stars decelerate and return to a bound state. To minimise these episodic fluctuations, we consider as ``escapers'' only stars that appear unbound for the last $5\,\Myr$ of the simulation (see the black lines in Fig.~\ref{fig: comparison methods unbound}, and also $\Mlost$ in Sect.~\ref{sec:outer_region}, Fig.~\ref{fig: mass_loss} and Tab.~\ref{tab:mass_loss}). The choice of this particular  interval is dictated by the time with which the autocorrelation of $\Xitid$, averaged between the simulations, crosses zero (see \citealt{suin_thesis} for further details). An alternative method is to label as escapers those stars that remain unbound for at least some predefined interval, $\tauw$, starting from the time of their first detection as unbound. In this way, the evolution of $\Mlost$ is  independent of the final state of the system.  A limitation is that its monotonicity might be lost if $\tauw$ were too small (see, e.g., the yellow line in Fig.~\ref{fig: comparison methods unbound}). If, however, $\tauw=5\,\Myr$ this method and the one we used in Sect.~\ref{sec:outer_region} are equivalent, with the latter being less computationally demanding.

\end{document}